\documentclass[fleqn,usenatbib]{mnras}

\usepackage{newtxtext,newtxmath}

\usepackage[T1]{fontenc}

\DeclareRobustCommand{\VAN}[3]{#2}
\let\VANthebibliography\thebibliography
\def\thebibliography{\DeclareRobustCommand{\VAN}[3]{##3}\VANthebibliography}

\usepackage{graphicx}	
\usepackage{amsmath}	

\usepackage{hyperref}
\hypersetup{colorlinks=true,allcolors=blue}

\usepackage{paralist}
\usepackage{lipsum}
\usepackage{siunitx}
\usepackage{mhchem}
\usepackage[]{subcaption}
\usepackage{rotating}
\usepackage{ulem}

\newcommand{\RM}[1]{\MakeUppercase{\romannumeral #1{}}}  		

\newcommand{\GlyOne}{Gly-\RM{1}}
\newcommand{\GlyOneS}{Gly-\RM{1} }
\newcommand{\GlyTwo}{Gly-\RM{2}}
\newcommand{\GlyTwoS}{Gly-\RM{2} }

\newcommand{\GlyForm}{\ce{NH2CH2COOH}}

\newcommand{\AceForm}{\ce{CH3CHO}}

\title[Deep Search for Glycine Conformers in Barnard 5]{Deep Search for Glycine Conformers in Barnard 5}

\author[T. Carl et al.]{
Tadeus Carl,$^{1}$\thanks{E-mail: tadeus.carl@chalmers.se}
E. S. Wirström,$^{1}$
P. Bergman,$^{1}$
S. B. Charnley,$^{2}$
Y.-L. Chuang$^{3}$ and
Y.-J. Kuan$^{3}$
\\
$^{1}$Chalmers University of Technology, Department of Space, Earth and Environment, Onsala Space Observatory, SE-439 92, Onsala, Sweden\\
$^{2}$Astrochemistry Laboratory and the Goddard Center for Astrobiology, Mailstop 691, 8800 Greenbelt Road, Greenbelt, MD 20770, USA\\
$^{3}$National Taiwan Normal University, Department of Earth Sciences, No. 88, Ting-Chou Rd., Wen-Shan District, Taipei City, Taiwan 11677, ROC
}

\date{Accepted 2023 June 30. Received 2023 June 30; in original form 2023 May 25}

\pubyear{2023}

\begin{document}
\label{firstpage}
\pagerange{\pageref{firstpage}--\pageref{lastpage}}
\maketitle

\begin{abstract}
One of the most fundamental hypotheses in astrochemistry and astrobiology states that crucial biotic molecules like glycine (\GlyForm) found in meteorites and comets are inherited from early phases of star formation. Most observational searches for glycine in the interstellar medium have focused on warm, high-mass molecular cloud sources. However, recent studies suggest that it might be appropriate to shift the observational focus to cold, low-mass sources. We aim to detect glycine towards the so-called methanol hotspot in the Barnard 5 dark cloud. The hotspot is a cold source ($T_\mathrm{gas}\approx 7.5$\,K) with yet high abundances of complex organic molecules (COMs) and water in the gas phase. We carried out deep, pointed observations with the Onsala 20\,m telescope, targeting several transitions of glycine conformers \RM{1} and \RM{2} (\GlyOneS and \GlyTwo) in the frequency range $70.2$\,--\,$77.9$\,GHz. No glycine lines are detected towards the targeted position, but we use a line stacking procedure to derive sensitive abundance upper limits w.r.t. \ce{H2} for \GlyOneS and \GlyTwo, i.e. $\leq(2$\,--\,$5)\times10^{-10}$ and $\leq(0.7$\,--\,$3)\times10^{-11}$, respectively. The obtained \GlyTwoS upper limits are the most stringent for a cold source, while the \GlyOneS upper limits are mostly on the same order as previously measured limits. The measured abundances w.r.t. \ce{H2} of other COMs at the B5 methanol hotspot range from $2\times10^{-10}$ (acetaldehyde) to $2\times10^{-8}$ (methanol). Hence, based on a total glycine upper limit of $(2$\,--\,$5)\times10^{-10}$, we cannot rule out that glycine is present but undetected.
\end{abstract}




\section{Introduction}

Glycine (\ce{NH2CH2COOH}) is the simplest amino acid used in the metabolism of terrestrial lifeforms. Since its first detection in meteorites in the 1970s \citep{Kvenvolden1970, CroninPizzarello1999, Ehrenfreund2001b, Botta2002, Glavin2006, KogaNaraoka2017} it has been speculated that prebiotic materials have been delivered to Earth by smaller impacting bodies \citep{HoyleWickramasinghe1977, EhrenfreundCharnley2000}. This hypothesis is supported by the identification of glycine in the comas of comets 81P/Wild\,2 \citep{Elsila2009} and 67P/Churyumov-Gerasimenko \citep{Altwegg2016, Hadraoui2019}, as well as in samples from asteroid (162173) Ryugu \citep{Naraoka2023}. There is general consensus that meteoric glycine is mainly the product of aqueous alteration in subsurface layers of the asteroidal host bodies \citep[][and references therein]{Lee2009, Ioppolo2021}. On the other hand, the study by \citet{Hadraoui2019} suggests that cometary glycine is more pristine and might indeed have an interstellar origin. 

To test the claimed hypothesis, astronomers have searched for signs of interstellar amino acids, in particular glycine, for more than four decades \citep{Brown1979, Hollis1980, Snyder1983, Berulis1985, GuelinCernicharo1989, Combes1996, Ceccarelli2000, Hollis2003a, Hollis2003b, Kuan2003, Snyder2005, Cunningham2007, Jones2007, Belloche2008, Guelin2008, Drozdovskaya2019, JimenezSerra2020}. However, no clear detection has been made to this day. Most studies have focused on high-mass star-forming regions (especially hot cores in Sgr\,B2 and Ori\,A), although some low-mass dark cloud sources have been included in the sample sets of \citet{Brown1979} and \citet{Snyder1983}. Only \citet{Ceccarelli2000} and \citet{Drozdovskaya2019} investigated low-mass protostellar sources (hot corinos). 

Hot cores with kinetic temperatures of 100\,--\,200\,K are often the first choice when searching for complex organic molecules (COMs) due to the overall large amount of available material combined with a rich chemistry and line variety.  In the case of glycine, the observational focus on hot cores has been justified by early theoretical \citep{Sorrell2001, Woon2002, Garrod2013} and experimental \citep{Bernstein2002, MunozCaro2002, Holtom2005, Lee2009} studies, indicating that glycine could only form by means of energetic (UV/electron/thermal) processing of interstellar ices, involving temperatures $>$\,50\,K. There is, however, increasing evidence that a shift of focus to cold ($\sim$\,10\,K) low-mass sources might be appropriate when searching for glycine. 

It is known from observations since the 1980s that COMs like methanol (\ce{CH3OH}), acetaldehyde (\ce{CH3CHO}), or methyl formate (\ce{CH3OCHO}) can reach significant abundances in cold molecular cloud sources such as TMC-1, L1689B, L1544, and Barnard 5 \citep{Matthews1985, Friberg1988, Marcelino2007, Bacmann2012, Vastel2014, JimenezSerra2016, Taquet2017}. Based on that high degree of chemical complexity, \citet{JimenezSerra2014} tested the detectability of glycine with radiative transfer models for the prototypical prestellar core L1544, indicating that several glycine transitions with an upper level energy $E_\mathrm{u} < 30$\,K and an Einstein coefficient of spontaneous emission $A_\mathrm{ul} \geq 10^{-6}$\,\si{\per\second} can reach detectable intensities in cold sources. Furthermore, severe observational issues with hot cores, such as line blending and multiple velocity components, can be avoided when observing colder, quiescent sources with lower line densities and narrower line widths. In addition, a combined experimental and theoretical study by \citet{Ioppolo2021} has recently shown for the first time that glycine can form on the surface of interstellar ices at low temperatures ($\sim$\,13\,K) and in the absence of energetic irradiation. Experiments and modelling agree that the main route of formation is the radical-radical reaction
\begin{equation}\label{eq:GlyForm1}
\ce{NH2CH2 + HOCO} \,\,\ce{->}\,\, \ce{NH2CH2COOH} .
\end{equation}

As many molecules, glycine has several stable conformational isomers with different zero-point energies. The two lowest-energy conformers of glycine are denoted with \RM{1} and \RM{2} (hereafter, \GlyOneS and \GlyTwo). The zero-point energy of \GlyTwoS is approximately $1007\pm101$\,K higher than that of the lower energy conformer \GlyOneS, while the electric dipole moment of \GlyTwoS is approximately six times higher than that of \GlyOneS \citep{Lovas1995}. Since spectral intensity is proportional to the square of the dipole moment, the intensity of \GlyTwoS is enhanced by a factor of approximately 36 over \GlyOne. This leads to an anomalous high intensity ratio of both conformers, i.e. a much weaker rotational spectrum of \GlyOneS compared to \GlyTwo, despite the lower zero-point energy of \GlyOneS \citep[cf.][]{Snyder1983}. For this reason, \GlyTwoS was the first conformer to be spectroscopically identified in the laboratory \citep{SuenramLovas1978, Brown1978, SuenramLovas1980}. It is also for this reason that early searches for glycine in the ISM \citep{Brown1979, Snyder1983, Berulis1985} focused on transitions of \GlyTwo, while all later studies focused either on \GlyOneS or included transitions of both conformers. 

\begin{figure}
\centering
\includegraphics[width=\hsize]{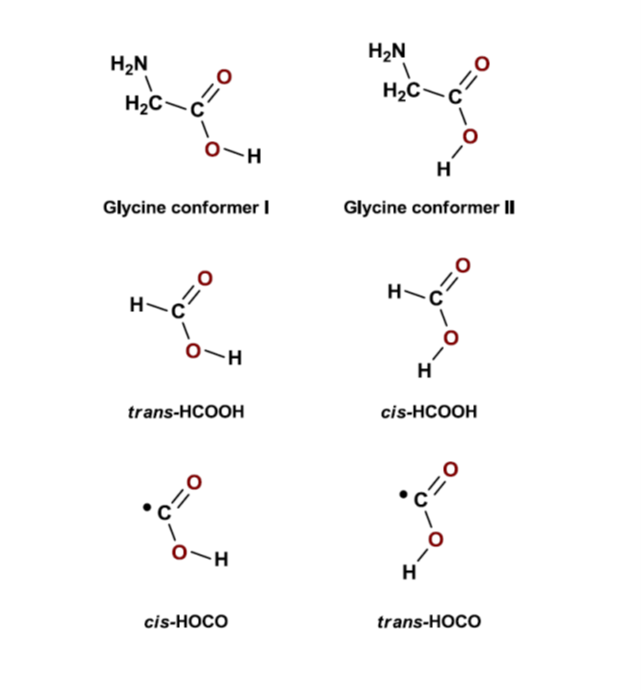}
  \caption{Schematic view of the conformers of parent and daughter molecules in the considered reactions that form glycine (\ce{NH2CH2COOH}) and formic acid (HCOOH) on an interstellar ice surface. Left column: hydrogenation of $cis$-HOCO leads to $trans$-HCOOH, while the radical-radical reaction \ce{NH2CH2 + }$cis$-HOCO leads to \GlyOne. Right column: hydrogenation of $trans$-HOCO leads to $cis$-HCOOH, and \ce{NH2CH2 + }$trans$-HOCO forms \GlyTwo. Credit: Jose Aponte.}
     \label{fig_GlyConfForm}
\end{figure}

Based on the spectroscopic properties of \GlyOneS and \GlyTwo, and on purely thermodynamic grounds, much higher abundances of \GlyOneS are expected over the whole molecular cloud temperature range (10\,--\,200\,K). However, observations have shown that an equilibrium isomerisation is not always established in molecular clouds, but rather that the high-energy conformer of a molecule can have much higher abundances than expected \citep[][]{Concepcion2022}. In the case of formic acid (HCOOH), the gas phase abundance ratio of the high-energy $c(is)$-conformer to the low-energy $t(rans)$-conformer can reach values of 5\,--\,6\%, as found towards two dark cloud sources, Barnard 5 and L483, with kinetic temperatures of around 10\,K \citep{Taquet2017, Agundez2019}. Those ratios are several tens of orders of magnitude higher than predicted by gas phase isomerisation models including ground-state quantum tunneling effects \citep{Concepcion2022}. This indicates that some key isomerisation mechanisms are not fully understood. Photo-switching mechanisms as proposed by \citet{Cuadrado2016} can likely be neglected in well-shielded dark cloud regions as B5 and L483. However, \citet{Taquet2017} suggest that the solid state formation of molecules on interstellar ices can produce non-equilibrium isomerisations because the outcome of a surface reaction depends also on the relative orientation between a surface species and the incoming gas phase species \citep[e.g.,][]{Rimola2012, Romero2016}. Furthermore, in the particular case of formic acid, it is assumed that one major route of solid state formation is the reaction chain
\begin{align}
\ce{CO + OH} &\,\,\ce{->}\,\, \ce{HOCO}      \\
\ce{HOCO + H} &\,\,\ce{->}\,\, \ce{HCOOH} \, , 
\end{align}    
with most of the HOCO radical formed in its low-energy $trans$-configuration \citep{Goumans2008}. However, the hydrogenation of $t$-HOCO leads to $c$-HCOOH, which can significantly increase the gas phase $cis$/$trans$ ratio of formic acid, once desorbed from the grain surface. We argue that the same principles can lead to anomalously high \GlyTwo/\GlyOneS gas phase ratios in cold sources like B5, if it is assumed that glycine forms in the solid phase via reaction (\ref{eq:GlyForm1}), and by noting that \GlyTwoS is formed from reactions of $t$-HOCO and \ce{NH2CH2} (cf. Fig.\,\ref{fig_GlyConfForm}). Howerver, recent modelling work by \citet{Garrod2022} suggests that several other glycine formation routes can be very efficient at dark cloud conditions and potentially dominate over reaction (\ref{eq:GlyForm1}). Of these, only the reaction of \ce{NH2} (or of NH followed by H) with \ce{CH2COOH} could lead to glycine conformers I and II since the latter radical is itself formed from HOCO. Thus, in this case, the \GlyTwo/\GlyOneS ratio should also follow the $c$/$t$-HCOOH. Therefore, gas phase observations of the \GlyTwo/\GlyOneS ratio in cold sources could in principle be used to investigate the main formation routes of glycine at low temperatures.


Based on the above considerations, the methanol hotspot\footnote{Following \citet{Wirstrom2014}, the term methanol "hotspot" is used to indicate the position of the methanol intensity peak.} in the Barnard 5 (hereafter B5) dark cloud \citep[Perseus, $d=302\pm21$\,pc; ][]{Zucker2018} is a promising target to search for interstellar glycine. It is located approximately 0.55\,pc to the northwest of the prominent class-\RM{1} protostar IRS-1 \citep[see maps in][]{Wirstrom2014, Taquet2017}. The location of the hotspot is neither associated with a prestellar core nor a protostellar source \citep[][and references therein]{Wirstrom2014}. The detection of abundant water ($8\times10^{-9}$) by \citet{Wirstrom2014} as well as methanol ($2\times10^{-8}$) and other COMs ($\sim$\,$10^{-10}$) by \citet{Taquet2017} confirms the chemical complexity of the region; abundances are relative to \ce{H2}. In addition, it is in this position that $c$-HCOOH has been detected at a relative abundance of 6\% as compared to $t$-HCOOH \citep{Taquet2017}. The \ce{H2} column density and dust temperature towards the hotspot are estimated from data of the Herschel Gould Belt survey\footnote{\url{http://www.herschel.fr/cea/gouldbelt/en/Phocea/Vie_des_labos/Ast/ast_visu.php?id_ast=66}} (HGBS) as $N(\ce{H2}) = 1\times10^{22}$\,\si{\per\square\centi\meter} and $T_\mathrm{dust} = 13$\,K, respectively\footnote{We used $N(\ce{H2}) = 1\times10^{22}$\,\si{\per\square\centi\meter} to update the abundances derived for water and COMs in \citet{Wirstrom2014} and \citet{Taquet2017}.} \citep[][]{Andre2010, Roy2014}. Non-LTE radiative transfer calculations of methanol emission by \citet{Taquet2017} reveal a gas kinetic temperature of $T_\mathrm{kin}= 7.5\pm1.5$\,K and a gas density of $n(\ce{H2}) = (2.25\pm1.50)\times10^{5}$\,\si{\per\cubic\centi\meter}. 

For our study, single-point observations were made with the Onsala 20\,m telescope towards the B5 hotspot, with several \GlyOneS and \GlyTwoS transitions targeted in the frequency range $\sim$\,$70$\,--\,$78$\,GHz. The observations are described in section \ref{sec_Observations}, and we present the resulting spectrum and the detected molecules in section \ref{sec_Results}. Glycine was not detected, but we derive sensitive upper limits for both glycine conformers using spectral line stacking, as described in section \ref{sec_Analysis}. In that section, we also present the column density calculation for other detected molecules. In section \ref{sec_Discussion}, the derived glycine upper limits are compared to limits for other sources as well as to reported abundances of other COMs at the position of the B5 hotspot. We summarise our conclusions in section \ref{sec_Conclusions}.

\section{Observations}\label{sec_Observations}

\begin{figure*}
\resizebox{\hsize}{!}
{\includegraphics[width=12cm,clip]{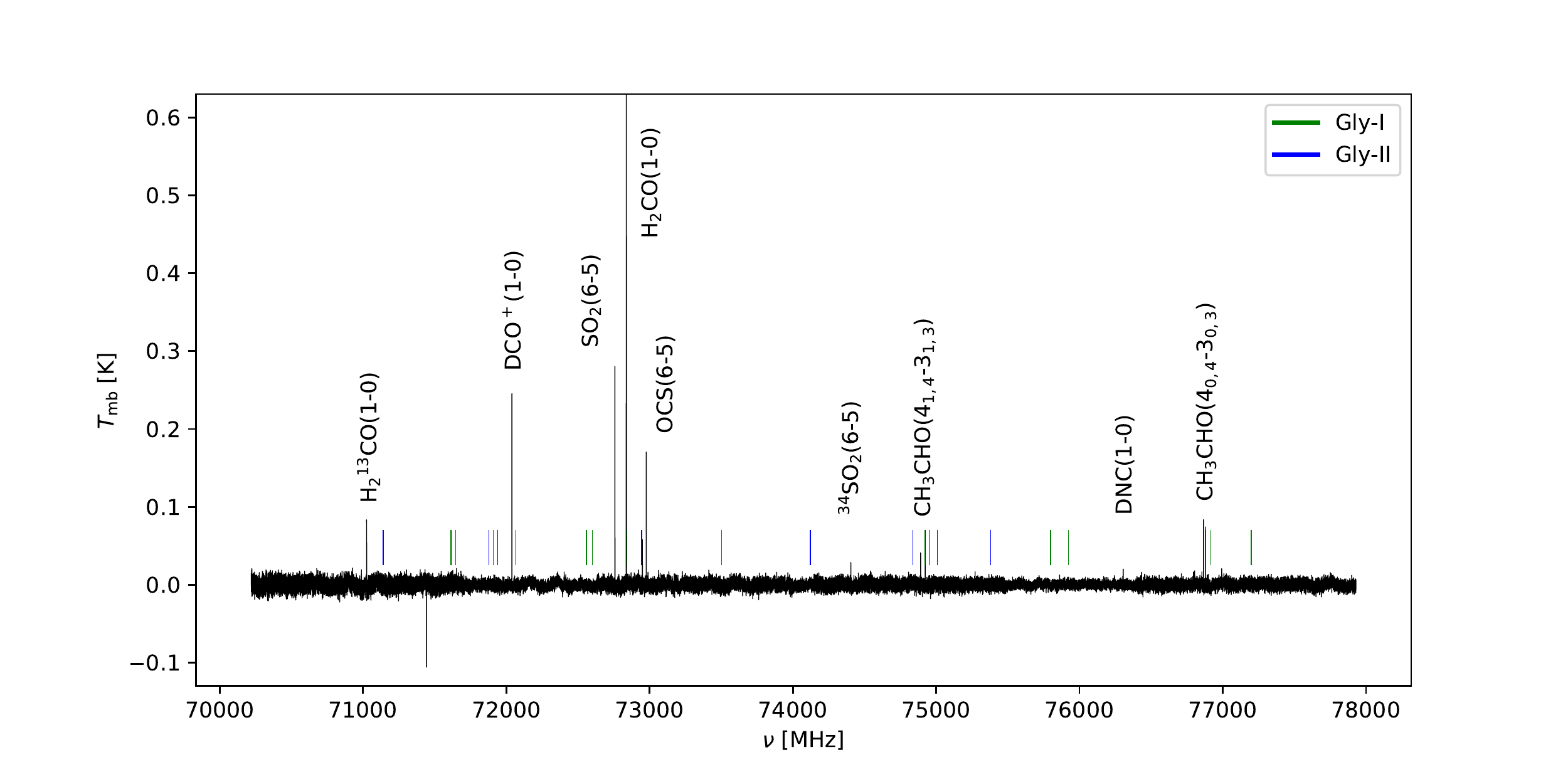}}
\caption{Spectrum towards the B5 methanol hotspot in the frequency range $70.2$\,--\,$77.9$\,GHz. The solid green and blue lines mark the frequencies of targeted \GlyOneS and \GlyTwoS transitions, respectively. The peak intensity of the truncated \ce{H2CO}(1-0) line is at $\sim$\,1.22\,K. The negative feature at $\sim$71.44\,GHz and the positive feature at $\sim$72.95\,GHz, left of the OCS(6-5) peak, are bad channels.}
\label{fig_Spectrum}
\end{figure*}

Our LTE analysis using $T_\mathrm{ex} = 7.5$\,K predicts that a number of \GlyOneS and \GlyTwoS transitions in the 4\,mm band will have detectable peak intensities if the glycine abundance and distribution at the B5 hotspot is similar to other detected COMs \citep{Taquet2017}. The considered transitions are presented in Table\,\ref{tab_GlyTrans} with their spectroscopic properties. Single point observations were carried out towards the B5 hotspot with the Onsala 20\,m telescope at $70.2$\,--\,$77.9$\,GHz. We observed for a total of $\sim$150\,h in three observation runs in 2017, 2018, and 2022. 

The receiver frontend was a 4\,mm dual-polarisation sideband-separating HEMT receiver \citep{Walker2016}. This frontend has common warm optics with the 3\,mm system \citep{Belitsky2015}. In the 2017 observations, the center frequency was 72\,GHz, and in 2018, it was 76\,GHz. In the 2022 observations, we used two settings with the center frequencies from 2017 and 2018, referred to as S1 and S2, respectively. We used a symmetric dual-beam switching mode with a switching frequency of 1\,Hz and a beam throw of 11$'$ in azimuth. The half-power beamwidth at 74\,GHz is 51$''$. We pointed the antenna towards 03:47:32.10\,RA (J2000) and 32:56:43.0\,Dec (J2000). Pointing and focus accuracy was checked with observations of SiO masers at $86.2$\,GHz, and was found to be within 5$''$ for the pointing. The receiver backend was a FFT spectrometer, consisting of four modules. To increase the effective observation time, we used the dual-polarisation mode, covering about 2.5\,GHz per module pair. This resulted in 4\,GHz total bandwidth per frequency setting, with an overlap of about 1\,GHz. The frequency resolution was 76.3\,kHz, corresponding to a velocity resolution of 0.31\,\si{\kilo\meter\per\second} at 74\,GHz. The data was calibrated using the chopper-wheel method and the estimated calibration uncertainty was 10\%. In the 2017, 2018, and 2022 observations, the system noise temperature varied mostly within 195\,--\,450\,K, 165\,--\,405\,K, and 170\,--\,360\,K, respectively, but for shorter periods of worse observing conditions it increased significantly above this. 

We used the 20\,m data reduction software XS for data reduction and correction for frequency- and elevation-dependent main beam efficiency and Doppler-shifted spectral resolution. Spectra with noise temperatures above 500\,K were excluded. We combined our 2017 and 2018 data with our 2022-S1 and 2022-S2 data, respectively, and averaged the dual-polarisation spectra weighted by their system temperature and integration time. We fitted and subtracted a linear baseline to line-free regions in the 2017-2022-S1 and 2018-2022-S2 averaged spectra, and combined them to one spectrum. 

\begin{table} 
\caption{\label{tab_GlyTrans}Targeted \GlyOneS and \GlyTwoS transitions and their spectroscopic properties, taken from CDMS \citep[\url{cdms.astro.uni-koeln.de}; see also][]{Ilyushin2005}.}
\centering
\begin{tabular}{ccccc}
\hline\hline
	Transition	&Frequency	&$E_\mathrm{u}$	&$A_\mathrm{ul}$	& $g_\mathrm{u}$		\rule{0pt}{2.5ex}	\\
				&[GHz]	&[K]		&[\si{\per\second}]		& [-]		\\
\hline
\GlyOne \rule{0pt}{2.5ex} \\
$10_{3,7}-9_{3,6}$	&	71.61156	&	21.5	&	$1.5\times10^{-6}$	&	63	\\
$11_{2,10}-10_{2,9}$	&	71.64639	&	22.4	&	$1.6\times10^{-6}$	&	69	\\
$10_{2,8}-9_{2,7}$	&	71.91030	&	20.2	&	$1.6\times10^{-6}$	&	63	\\
$12_{1,12}-11_{1,11}$	&	72.55935	&	23.3	&	$1.8\times10^{-6}$	&	75	\\
$12_{0,12}-11_{0,11}$	&	72.60111	&	23.3	&	$1.8\times10^{-6}$	&	75	\\
$11_{1,10}-10_{1,9}$	&	72.84125	&	22.3	&	$1.7\times10^{-6}$	&	69	\\
$6_{3,4}-5_{2,3}$		&	73.50346	&	9.9	&	$3.5\times10^{-7}$	&	39	\\
$11_{3,9}-10_{3,8}$	&	74.92344	&	24.7	&	$1.8\times10^{-6}$	&	69	\\
$4_{4,1}-3_{3,0}$		&	75.79527	&	8.6	&	$7.2\times10^{-7}$	&	27	\\
$4_{4,0}-3_{3,1}$		&	75.80214	&	8.6	&	$7.2\times10^{-7}$	&	27	\\
$11_{4,8}-10_{4,7}$	&	75.92223	&	27.1	&	$1.8\times10^{-6}$	&	69	\\
$11_{4,7}-10_{4,6}$	&	76.91370	&	27.2	&	$1.8\times10^{-6}$	&	69	\\
$6_{3,3}-5_{2,4}$		&	77.19906	&	9.9	&	$3.7\times10^{-7}$	&	39	\\
$12_{2,11}-11_{2,10}$	&	77.65229	&	26.1	&	$2.1\times10^{-6}$	&	75	\\
\hline
\GlyTwo \rule{0pt}{2.5ex} \\									
$10_{3,8}-9_{3,7}$	&	71.14150	&	21.7	&	$5.2\times10^{-5}$	&	63	\\
$10_{6,5}-9_{6,4}$	&	71.61483	&	30.2	&	$3.8\times10^{-5}$	&	63	\\
$10_{6,4}-9_{6,3}$	&	71.61729	&	30.2	&	$3.8\times10^{-5}$	&	63	\\
$10_{5,6}-9_{5,5}$	&	71.87704	&	26.8	&	$4.5\times10^{-5}$	&	63	\\
$10_{5,5}-9_{5,4}$	&	71.94225	&	26.8	&	$4.5\times10^{-5}$	&	63	\\
$10_{4,7}-9_{4,6}$	&	72.06595	&	24.0	&	$5.0\times10^{-5}$	&	63	\\
$10_{4,6}-9_{4,5}$	&	72.94452	&	24.0	&	$5.2\times10^{-5}$	&	63	\\
$11_{2,10}-10_{2,9}$	&	74.12176	&	23.1	&	$6.2\times10^{-5}$	&	69	\\
$10_{2,8}-9_{2,7}$	&	74.83870	&	21.0	&	$6.4\times10^{-5}$	&	63	\\
$12_{1,12}-11_{1,11}$	&	74.92670	&	24.1	&	$6.7\times10^{-5}$	&	75	\\
$12_{0,12}-11_{0,11}$	&	74.95073	&	24.1	&	$6.7\times10^{-5}$	&	75	\\
$11_{1,10}-10_{1,9}$	&	75.00964	&	23.0	&	$6.5\times10^{-5}$	&	69	\\
$10_{3,7}-9_{3,6}$	&	75.37886	&	22.2	&	$6.3\times10^{-5}$	&	63	\\
$11_{3,9}-10_{3,8}$	&	77.91634	&	25.4	&	$7.0\times10^{-5}$	&	69	\\

\hline
\end{tabular}
\end{table}

\section{Resulting Spectrum and Line Identification}\label{sec_Results}

The resulting spectrum of our observations towards the B5 hotspot is shown in Fig.\,\ref{fig_Spectrum}. The frequencies of the targeted \GlyOneS and \GlyTwoS transitions from Table\,\ref{tab_GlyTrans} are marked by solid green and blue lines, respectively. The selection of considered glycine transitions is described in section \ref{sec_ULGly}. The RMS noise towards the low-frequency end of the spectrum is $\sim$\,6.2\,mK. It decreases with frequency to reach  $\sim$\,4.2\,mK towards the high-frequency end. In regions where the backend modules overlap, around 72\,GHz and 76\,GHz, there is a noticeable decrease in noise due to the spectral resampling resulting in an increase of the effective channel width.

None of the targeted glycine lines shows an intensity above the noise level but we use spectral line stacking to derive column density and abundance upper limits for \GlyOneS and \GlyTwo, separately. This is described in more detail in section \ref{sec_ULGly}. 

We do detect emission lines from acetaldehyde (\ce{CH3CHO}), formaldehyde (\ce{H2CO}; H$_2$\ce{^{13}CO}), the \ce{DCO^+} ion, deuterated hydrogen isocyanide (DNC), carbonyl sulfide (OCS), and sulfur dioxide (\ce{SO2}; \ce{^{34}SO2}). The detected transitions are listed in column two of Table\,\ref{tab_Detections} together with their line width and integrated intensity (last two columns). The individual spectral lines with their Gaussian fits are shown in Fig.\,\ref{fig_Spectra}. 

\section{Analysis}\label{sec_Analysis}

\begin{table*} 
\caption{\label{tab_Detections}Detected transitions with their spectroscopic and observed parameters; all uncertainties are 1-sigma values. Spectroscopic parameters are taken from CDMS (\url{cdms.astro.uni-koeln.de}) and JPL \citep[\url{spec.jpl.nasa.gov}; only for \ce{CH3CHO}; see also][]{Kleiner1996}.}
\centering
\begin{tabular}{lcccccccc}
\hline\hline
Molecule 	&Transition	&$\nu$	&$E_\mathrm{u}$	&$A_\mathrm{ul}$	&$g_\mathrm{u}$	&$Q_\mathrm{rot}(9.375\,\mathrm{K})$ 	&FWHM$^\mathrm{(a)}$	&$\int T_\mathrm{mb}\mathrm{d}v^\mathrm{(b)}$  \rule{0pt}{2.5ex} \\
		&		&[GHz]	&[K]			&[\si{\per\second}]	& 			&		&[\si{\kilo\meter\per\second}]&[\si{\milli\kelvin\kilo\meter\per\second}]	\\
\hline

\ce{A-CH3CHO}	&	$4_{1,4}-3_{1,3}$	&	74.89167	&	11.3	&	$1.3\times10^{-5}$	&	9$^\mathrm{(c)}$	&	66.02$^\mathrm{(d)}$	&	$0.67	\pm	0.07$	&	$30.2	\pm	3.3$	\\
	&	$4_{0,4}-3_{0,3}$	&	76.87895	&	9.2	&	$1.5\times10^{-5}$	&	9$^\mathrm{(c)}$	&	66.02$^\mathrm{(d)}$	&	$0.51	\pm	0.06$	&	$45.8	\pm	2.8$	\\
\ce{E-CH3CHO}	&	$4_{1,4}-3_{1,3}$	&	74.92413	&	11.3	&	$1.3\times10^{-5}$	&	9$^\mathrm{(c)}$	&	66.02$^\mathrm{(d)}$	&	$0.75	\pm	0.10$	&	$27.0	\pm	2.7$	\\
	&	$4_{0,4}-3_{0,3}$	&	76.86643	&	9.3	&	$1.5\times10^{-5}$	&	9$^\mathrm{(c)}$	&	66.02$^\mathrm{(d)}$	&	$0.67	\pm	0.03$	&	$60.8	\pm	2.9$	\\
\ce{H2CO}	&	$1_{0,1}-0_{0,0}$	&	72.83794	&	3.5	&	$8.2\times10^{-6}$	&	3	&	13.80	&	$1.04	\pm	0.04$	&	$1370.1	\pm	3.8$	\\
H$_2$\ce{^{13}CO}	&	$1_{0,1}-0_{0,0}$	&	71.02478	&	3.4	&	$7.6\times10^{-6}$	&	3	&	14.13	&	$0.77	\pm	0.07$	&	$69.3	\pm	4.9$	\\
\ce{DCO^+}	&	$1-0$	&	72.03931	&	3.5	&	$2.3\times10^{-5}$	&	3	&	5.77	&	$1.29	\pm	0.06$	&	$331.6	\pm	3.0$	\\
DNC	&	$1-0$	&	76.30573	&	3.7	&	$1.6\times10^{-5}$	&	3	&	5.47	&	$1.92	\pm	0.27$	&	$29.1	\pm	2.4$	\\
OCS	&	$6-5$	&	72.97677	&	12.3	&	$1.1\times10^{-6}$	&	13	&	32.46	&	$0.60	\pm	0.03$	&	$102.2	\pm	3.5$	\\
\ce{SO2}	&	$6_{0,6}-5_{1,5}$	&	72.75824	&	19.2	&	$2.8\times10^{-6}$	&	13	&	33.07	&	$0.70	\pm	0.02$	&	$220.5	\pm	4.0$	\\
\ce{^{34}SO2}	&	$6_{0,6}-5_{1,5}$	&	74.40457	&	19.1	&	$3.0\times10^{-6}$	&	13	&	33.64	&	$0.49	\pm	0.13$	&	$14.0	\pm	2.6$	\\

\hline
\multicolumn{9}{l}{\small{$^\mathrm{(a)}$ derived from Gaussian fits (see Fig.\,\ref{fig_Spectra}); $^\mathrm{(b)}$ derived by integration over line channels (see Fig.\,\ref{fig_Spectra}).}} \rule{0pt}{2.5ex} \\
\multicolumn{9}{l}{\small{$^\mathrm{(c)}$ we use $g_\mathrm{u} = 2J+1$ to be consistent with $Q_\mathrm{rot}(T)$ from \citet{Nummelin1998}.}} \\
\multicolumn{9}{l}{\small{$^\mathrm{(d)}$ from \citet{Nummelin1998} for A-/E-species separated, i.e. $Q_\mathrm{rot}(T) = 2.3\times T^{3/2}$.}} 

\end{tabular}
\end{table*}

\begin{table}
\caption{\label{tab_CDAbun}Beam-averaged column densities and abundances relative to \ce{H2}; uncertainties are 1-sigma values and upper limits are 3-sigma values. We use $N(\ce{H2}) = 1\times10^{22}$\,\si{\per\square\centi\meter} from HGBS data \citep{Roy2014}.} 
\centering
\begin{tabular}{lcc}
\hline\hline
Molecule	& $N_\mathrm{tot}$	& $N_\mathrm{tot}/N(\ce{H2})$ 	\rule{0pt}{2.5ex} \\	
		& [\si{\per\square\centi\meter}]		& 			\\
\hline
\AceForm$^\mathrm{(a)}$			&	$(1.97\pm0.65)\times10^{12}$	&	$2\times10^{-10}$	\rule{0pt}{2.5ex}\\
\ce{H2CO}	&	$(4.14\pm0.01)\times10^{13}$	&	$4\times10^{-9}$	\\
H$_2$\ce{^{13}CO}	&	$(6.20\pm0.44)\times10^{11}$	&	$6\times10^{-11}$	\\
\ce{DCO^+}	&	$(4.14\pm0.04)\times10^{11}$	&	$4\times10^{-11}$	\\
DNC	&	$(6.27\pm0.51)\times10^{10}$	&	$6\times10^{-12}$	\\
OCS	&	$(1.31\pm0.04)\times10^{13}$	&	$1\times10^{-9}$	\\
\ce{SO2}	&	$(2.48\pm0.04)\times10^{13}$	&	$3\times10^{-9}$	\\
\ce{^{34}SO2}	&	$(1.54\pm0.28)\times10^{12}$	&	$2\times10^{-10}$	\\
\\		
\GlyOne	&	$\leq4.6\times10^{12}$$^\mathrm{(b)}$	&	$\leq5\times10^{-10}$$^\mathrm{(b)}$		\\
		&	$\leq2.1\times10^{12}$$^\mathrm{(c)}$	&	$\leq2\times10^{-10}$$^\mathrm{(c)}$	\\
        &	$\leq1.6\times10^{12}$$^\mathrm{(d)}$	&	$\leq2\times10^{-10}$$^\mathrm{(d)}$	\\
\\
\GlyTwo	&	$\leq3.3\times10^{11}$$^\mathrm{(b)}$	&	$\leq3\times10^{-11}$$^\mathrm{(b)}$	\\
		&	$\leq1.1\times10^{11}$$^\mathrm{(c)}$	&	$\leq1\times10^{-11}$$^\mathrm{(c)}$	\\
	    &	$\leq7.1\times10^{10}$$^\mathrm{(d)}$	&	$\leq7\times10^{-12}$$^\mathrm{(d)}$	\\

\hline

\multicolumn{3}{l}{$^\mathrm{(a)}$\small{\,$T_\mathrm{ex} = 5.7$\,K, from \ce{CH3CHO} rotation diagram analysis.}} \rule{0pt}{2.5ex} \\
\multicolumn{3}{l}{$^\mathrm{(b)}$\small{\,$T_\mathrm{ex} = 5$\,K;} $^\mathrm{(c)}$\small{\,$T_\mathrm{ex} = 7.5$\,K;} $^\mathrm{(d)}$\small{\,$T_\mathrm{ex} = 10$\,K.}} \\

\end{tabular}
\end{table}

\subsection{Glycine Upper Limits}\label{sec_ULGly}

Spectral line stacking is used to derive the highest sensitivity estimates possible for the column density upper limits of \GlyOneS and \GlyTwo. Our stacking method consists of four steps: 
\begin{enumerate}[(1)]
\item identification of the presumably strongest transitions, 
\item extraction of spectra around the transition frequencies of the identified transitions, 
\item conversion of intensity scale to column density scale in all spectra, assuming LTE conditions, 
\item averaging of the obtained column density spectra. 
\end{enumerate}

Assuming LTE conditions and optically thin emission, the total, beam-averaged column density of a molecule can be written as
\begin{equation}\label{eq:CD}
N_\mathrm{tot} = \frac{1}{C}\int T_\mathrm{mb}\mathrm{d}v\, , 
\end{equation}
where $\int T_\mathrm{mb}\mathrm{d}v$ is the integrated intensity and  $C$ is a constant, given as
\begin{equation}\label{eq:C}
C = \frac{hc^3}{8\pi k_\mathrm{B}} \frac{g_\mathrm{u}A_\mathrm{ul}}{\nu^2} \frac{\exp(-E_\mathrm{u}/T_\mathrm{ex})}{Q_\mathrm{rot}(T_\mathrm{ex})} \Biggl[1 - \frac{J_\nu(T_\mathrm{bg})}{J_\nu(T_\mathrm{ex})}\Biggr]\, .
\end{equation}
$h$ is the Planck constant, $c$ is the speed of light, $k_\mathrm{B}$ is the Boltzmann constant, $T_\mathrm{ex}$ is the excitation temperature, and $T_\mathrm{bg}$ is the cosmic microwave background (CMB) temperature. The line-specific spectroscopic parameters are the total degeneracy $g_\mathrm{u}$, Einstein coefficient for spontaneous emission $A_\mathrm{ul}$, transition (rest) frequency $\nu$, upper state energy $E_\mathrm{u}$ (in Kelvin), and the rotational partition function $Q_\mathrm{rot}$ at $T = T_\mathrm{ex}$. Furthermore, $J_\nu(T)$ is the temperature equivalent of the specific intensity, given as
\begin{equation}
J_\nu(T) = \frac{h\nu}{k_\mathrm{B}}\Biggl[\exp\Biggl(\frac{h\nu}{k_\mathrm{B} T}\Biggr) - 1\Biggr]^{-1} \, .
\end{equation}
The last term in Eq.\,(\ref{eq:C}) is only needed if $T_\mathrm{ex}$ is comparable to $T_\mathrm{bg}$, which is the case for the B5 hotspot. According to Eq.\,(\ref{eq:CD}), the column density in one spectral channel of width $\Delta v$ is given by
\begin{equation}
N_\mathrm{tot}^\mathrm{ch} = \frac{T_\mathrm{A}^\mathrm{ch}\Delta v }{C}\, ,
\end{equation}
and for step (1), we estimate the antenna temperature in one channel ($T_\mathrm{A}^\mathrm{ch}$) for all \GlyOneS and \GlyTwoS transitions in the frequency range $70.2$\,--\,$77.9$\,GHz. We assume $N_\mathrm{tot}^\mathrm{ch} = 3.6\times10^{12}$\,\si{\per\square\centi\meter} for \GlyOneS and $N_\mathrm{tot}^\mathrm{ch} = 2.0\times10^{11}$\,\si{\per\square\centi\meter} for \GlyTwo, and set $\Delta v = 0.4$\,\si{\kilo\meter\per\second} and $T_\mathrm{ex} = 7.5$\,K, based on temperature estimates for the B5 methanol hotspot \citep{Taquet2017}. Only transitions for which $T_\mathrm{A}^\mathrm{ch} > T_\mathrm{lim}$ are considered, and we use $2$\,mK and $1$\,mK as temperature limits $T_\mathrm{lim}$ for \GlyOneS and \GlyTwo, respectively. The considered transitions are listed in Table\,\ref{tab_GlyTrans}, together with the spectroscopic parameters that feed into the constant $C$, i.e. $E_\mathrm{u}$, $A_\mathrm{ul}$, and $g_\mathrm{u}$ \citep{Ilyushin2005}. The rotational partition functions of \GlyOneS and \GlyTwoS are 1350.38 and 1310.26 at $T = 9.375$\,K, respectively (values from CDMS\footnote{\url{cdms.astro.uni-koeln.de}}). Based on the theoretical temperature dependence for the rotational partition function of (a)symmetric-top molecules, we determine the partition function at $T_\mathrm{ex}$ as
\begin{equation}\label{eq:QrotSymTopTransform}
Q_\mathrm{rot}(T_\mathrm{ex}) = Q_\mathrm{rot}(T)\Biggl(\frac{T_\mathrm{ex}}{T}\Biggr)^{3/2} \, .
\end{equation} 

As step (2), we extract spectra over 50\,\si{\kilo\meter\per\second} around the transition frequencies of the considered \GlyOneS and \GlyTwoS transitions, and determine 1-sigma noise temperatures $T_\mathrm{mb}^\sigma$ in line-free regions (compare Fig.\,\ref{fig_Gly1Spectra} and Fig.\,\ref{fig_Gly2Spectra}). In order to obtain a single column density estimate from the obtained spectra, the individual spectroscopic line parameters must be included before averaging. In step (3), we do this by transferring the intensity axis of each spectrum to a column density axis by using 
\begin{equation}\label{eq:CDscale}
N_\mathrm{tot}(v) = \frac{T_\mathrm{mb}(v)\Delta v }{C}\, ,
\end{equation}
where $T_\mathrm{mb}(v)$ is the observed main beam temperature. Finally, as step (4), we average the obtained column density spectra weighted by 1-sigma noise column density, which we calculate for all spectra as
\begin{equation}\label{eq:1sigmaCD_indSpec}
N_\mathrm{tot}^\sigma = \frac{T_\mathrm{mb}^\sigma\Delta v}{C}\, .
\end{equation}

\begin{figure*}
\resizebox{\hsize}{!}
{\includegraphics[width=12cm,clip]{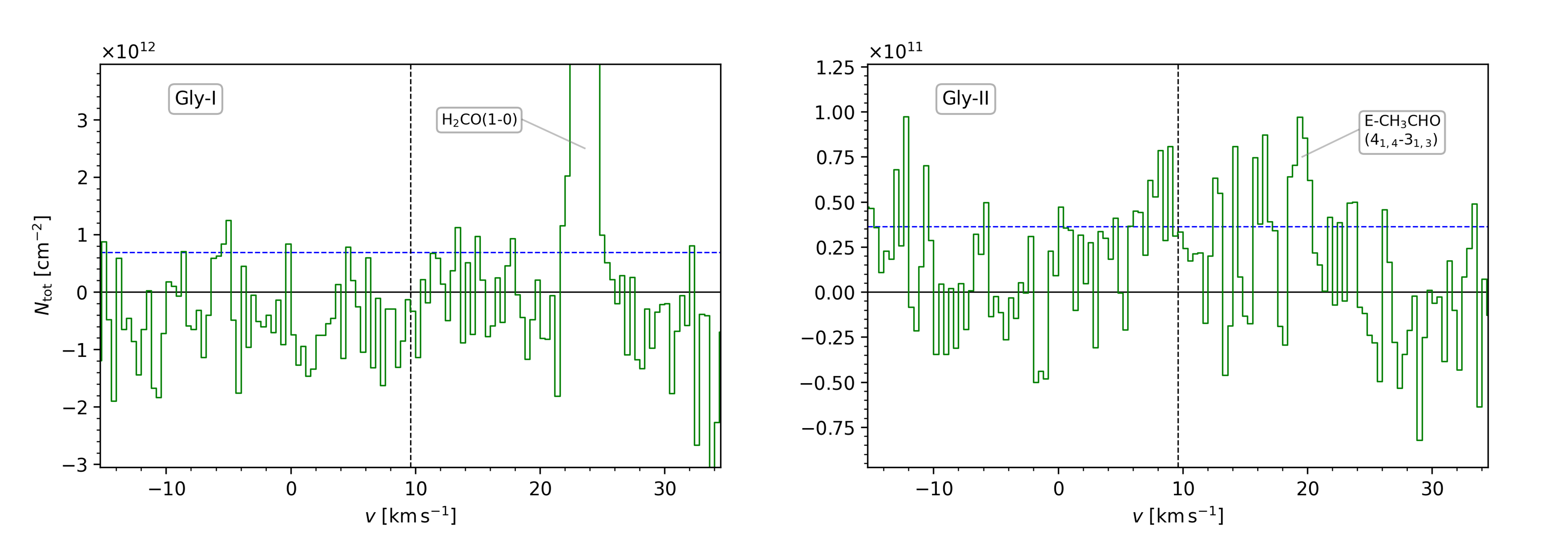}}
\caption{Stacked spectra for \GlyOneS (left) and \GlyTwoS (right) in column density scale, determined from Eq.\,(\ref{eq:CDscale}). In both spectra, the dashed blue line marks the 1-sigma column density noise level, and the dashed black line marks the LSR velocity of the B5 methanol hotspot. The column density peaks at $\sim$\,24\,\si{\kilo\meter\per\second} (left) and $\sim$\,20\,\si{\kilo\meter\per\second} (right) arise from the detected \ce{H2CO}(1-0) and \ce{CH3CHO}($4_{1,4}$-$3_{1,3}$) transitions, respectively. The regions where the peaks appear are excluded from the RMS calculation (compare Fig.\,\ref{fig_Gly1Spectra} and \ref{fig_Gly2Spectra}).}
\label{fig_CDSpectra}
\end{figure*}

The stacked column density spectra for \GlyOneS and \GlyTwoS are shown in Fig.\,\ref{fig_CDSpectra}. Even in the stacked spectra, glycine is still not detected, but we use the obtained noise level to derive sensitive upper limits. In both spectra, the dashed blue line marks the 1-sigma column density noise level. The corresponding 3-sigma column density upper limits of \GlyOneS and \GlyTwoS are presented in Table\,\ref{tab_CDAbun} together with the abundance upper limits relative to \ce{H2}.

\subsection{Column Densities of Detected Species}\label{sec_CDDet}

Column densities of \ce{H2CO}, H$_2$\ce{^{13}CO}, \ce{DCO^+}, DNC, OCS, \ce{SO2}, and \ce{^{34}SO2} are calculated from Eq.\,(\ref{eq:CD}), assuming LTE conditions and optically thin emission. The required spectroscopic parameters are given in Table\,\ref{tab_Detections}, and we use Eq.\,(\ref{eq:QrotSymTopTransform}) to determine the rotational partition functions at $T_\mathrm{ex} = 7.5$\,K. For the linear molecules DNC and OCS, we use the equivalent relation 
\begin{equation}\label{eq:QrotLinearTransform}
Q_\mathrm{rot}(T_\mathrm{ex}) = Q_\mathrm{rot}(T)\frac{T_\mathrm{ex}}{T} \, .
\end{equation} 

In the case of the \ce{H2CO}(1-0) line, we expect the optically thin approximation to fail, and derive an estimate for the optical depth $\tau$ using the rare isotopologue method with
\begin{equation}
\frac{W_{12}}{W_{13}} = [1 - \exp(-\tau)] \Biggl[1 - \exp\Biggl(\frac{-\tau}{\ce{^{12}C}/\ce{^{13}C}}\Biggr)\Biggr]^{-1}\, ,
\end{equation}
where $W_{12}$ and $W_{13}$ are the integrated intensities of \ce{H2CO}(1-0) and H$_2$\ce{^{13}CO}(1-0), respectively. We assume $\ce{^{12}C}/\ce{^{13}C} = 68$ for the local ISM, based on \citet{Milam2005}. We get $\tau = 3.4$ and calculate the corrected column density of \ce{H2CO} using 
\begin{equation}
N_\mathrm{tot}^\mathrm{corr} = \frac{\tau}{1 - \exp(-\tau)} N_\mathrm{tot}^\mathrm{thin}\, ,
\end{equation}
where $N_\mathrm{tot}^\mathrm{thin}$ is the optically thin estimate. 

The derived column densities are presented in Table\,\ref{tab_CDAbun}, together with the molecular abundances relative to \ce{H2}.

\subsection{Rotation Diagram Acetaldehyde}\label{sec_RDAce}

\begin{figure}
\centering
\includegraphics[width=\hsize]{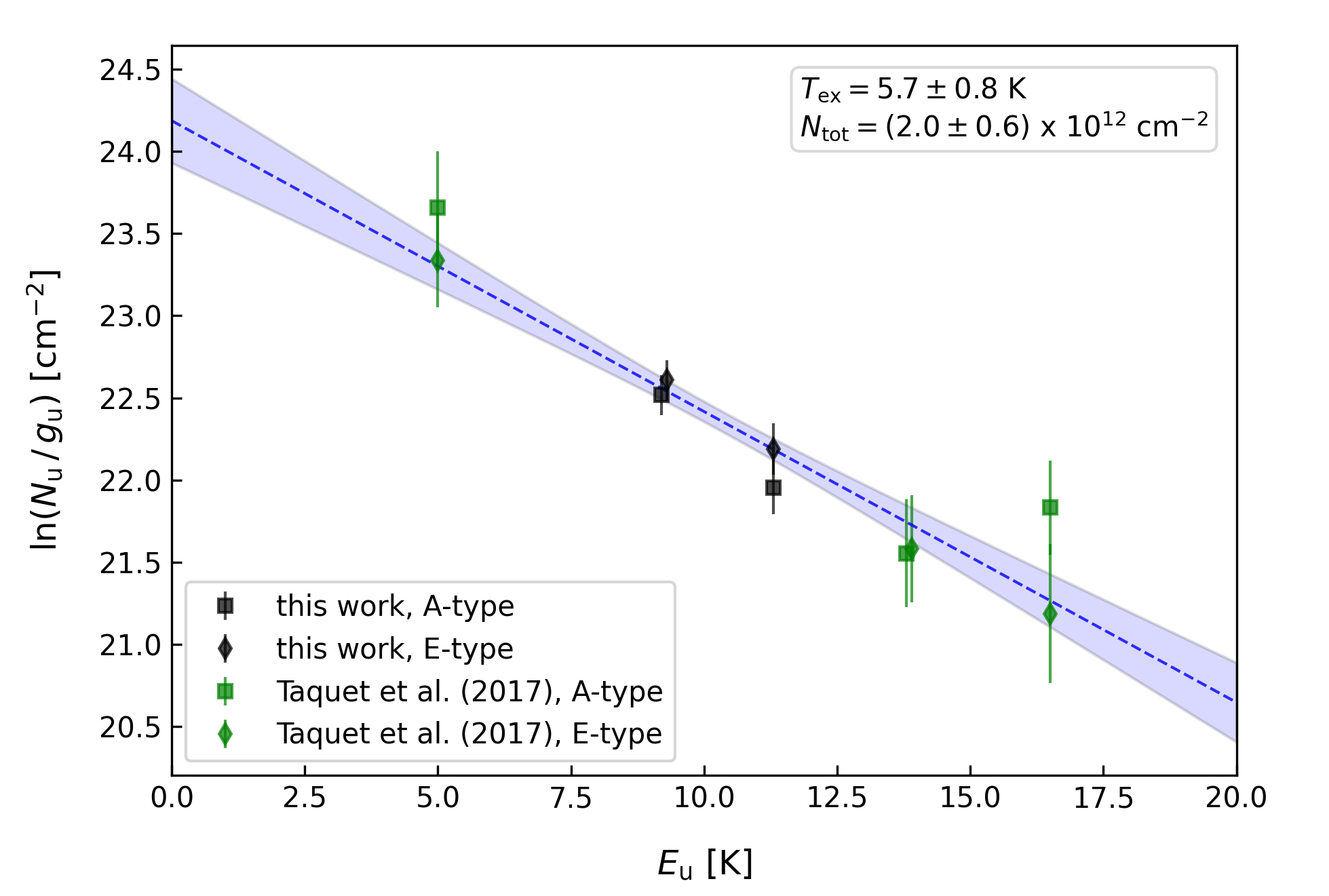}
\caption{Rotation diagram for acetaldehyde (\ce{CH3CHO}) transitions detected in this work (black symbols) and in \citet{Taquet2017} (green symbols) towards the B5 methanol hotspot. A-type and E-type transitions are depicted as squares and diamonds, respectively. The best linear fit to the data is shown as dashed blue line with 1-sigma confidence interval (shaded blue area). Considering the influence of CMB radiation, the resulting excitation temperature is $5.7\pm0.8$\,K, and the total column density is $(2.0\pm0.6)\times10^{12}$\,\si{\per\square\centi\meter}.}
\label{fig_RotD}
\end{figure}

Following \citet{GoldsmithLanger1999}, we perform a rotation diagram analysis for the four detected transitions of acetaldehyde (\ce{CH3CHO}), listed in Table\,\ref{tab_Detections}, together with six transitions detected towards the B5 hotspot by \citet{Taquet2017}. We assume optically thin emission and an LTE population of levels, and calculate the upper state column densities $N_\mathrm{u}$ as
\begin{equation}\label{eq:USCD}
N_\mathrm{u} = \frac{8\pi k_\mathrm{B}\nu^2}{A_\mathrm{ul}hc^3}\int T_\mathrm{mb}\mathrm{d}v\, .
\end{equation}
Plotting $\ln(N_\mathrm{u}/g_\mathrm{u})$ against $E_\mathrm{u}$ and fitting the data linearly yields an excitation temperature of $T_\mathrm{ex} = 5.8 \pm 0.8$\,K and a total column density of $N_\mathrm{tot} = (1.4\pm0.5)\times10^{12}$\,\si{\per\square\centi\meter}. Fitting the data for A- and E-species separately yields approximately the same result as fitting the data for both species combined, as expected from the low difference in ground state energy \citep[$0.1$\,K,][]{Kleiner1996}. When finding an excitation temperature so close to the CMB temperature ($T_\mathrm{bg}=2.73$\,K) it is important to consider its influence on excitation using the correction term 
\begin{equation}\label{eq:Cbg}
C_\mathrm{bg} = \Biggl[1 - \frac{J_\nu(T_\mathrm{bg})}{J_\nu(T_\mathrm{ex})}\Biggr]^{-1} \, .
\end{equation}
The corrected upper state column density is then given by
\begin{equation}
N_\mathrm{u}^\mathrm{corr} = C_\mathrm{bg}\,N_\mathrm{u} \, .
\end{equation}

Iterating the updated values of $N_\mathrm{u}^\mathrm{corr}$ and $T_\mathrm{ex}$, the asymptotic values of excitation temperature and total column density are $T_\mathrm{ex} = 5.7$\,K and $N_\mathrm{tot} = (2.0\pm0.6)\times10^{12}$\,\si{\per\square\centi\meter}. The resulting abundance relative to \ce{H2} is $2\times10^{-10}$ (see Table\,\ref{tab_CDAbun}). The final rotation diagram is shown in Fig.\,\ref{fig_RotD}. For the calculation of total column density, we use the rotational partition function derived in \citet{Nummelin1998} as $Q_\mathrm{rot}(T) = 4.6\times T^{3/2}$ for A- and E-species combined, and for $g_\mathrm{u} = 2J + 1$. Considering both species separately using the same iterative process as described gives an A/E ratio of $0.88$, which is close to the expected ratio of unity. 

\section{Discussion}\label{sec_Discussion}

The excitation temperature of $5.7\pm0.8$\,K obtained from the \ce{CH3CHO} data indicates possible sub-thermal excitation of the rotational levels of acetaldehyde at the position of the B5 hotspot (where the gas kinetic temperature is estimated from methanol lines as $T_\mathrm{kin}=7.5\pm1.5$\,K). Based on that, and in order to cover a larger range of possible excitation temperatures for glycine, we calculate two more sets of 3-sigma upper limits for $5$\,K and $10$\,K (see Table\,\ref{tab_CDAbun}). For the lower excitation limit of $T_\mathrm{ex} = 5$\,K, the number of considered \GlyOneS and \GlyTwoS transitions in the spectral line stacking reduces to 2 and 10, respectively, and the abundance upper limits increase to  $5\times10^{-10}$ and $3\times10^{-11}$. Conversely, for the higher excitation limit of $T_\mathrm{ex} = 10$\,K, the number of considered \GlyOneS and \GlyTwoS transitions increases to 18 and 16, respectively. The abundance upper limit of \GlyOneS remains at rounded $2\times10^{-10}$, while that of \GlyTwoS decreases to $7\times10^{-12}$. 

\subsection{Glycine Upper Limits compared to other Sources}

Few previous attempts have been made to detect glycine towards cold low mass sources. The published upper limits we have identified are summarised in Table\,\ref{tab_ULcomparison}. 3-sigma upper limits\footnote{Reported are 2-sigma limits from which we derived 3-sigma values.} of \GlyTwoS are reported for the dark clouds TMC-1 and L183 in \citet{Brown1979} as $3\times10^{12}$\,\si{\per\square\centi\meter} and $4.5\times10^{12}$\,\si{\per\square\centi\meter}, respectively, assuming $T_\mathrm{ex} = 7$\,K for both sources. Another 3-sigma upper limit of \GlyTwoS is reported in \citet{Snyder1983} for L134 as $2\times10^{12}$\,\si{\per\square\centi\meter}, assuming $T_\mathrm{ex} = 10$\,K. Adopting the respective \ce{H2} column densities towards TMC-1, L183, and L134 \citep[$2\times10^{22}$\,\si{\per\square\centi\meter}, $3\times10^{22}$\,\si{\per\square\centi\meter}, $1\times10^{22}$\,\si{\per\square\centi\meter};][]{Spezzano2022, Lattanzi2020, vanderWerf1988}, the corresponding \GlyTwoS abundance upper limits are presented in Table\,\ref{tab_ULcomparison}. Depending on the assumed excitation temperature at the B5 hotspot, the \GlyTwoS abundance upper limit of this work is one to two orders of magnitude lower than those early estimates. 

\citet{JimenezSerra2016} report two sets of 3-sigma upper limits for \GlyOneS towards the centre of the prestellar core L1544 and an offset position towards the shell-region where methanol emission is found to peak (see Table\,\ref{tab_ULcomparison}). The assumed excitation temperatures are 5\,K and 10\,K. Based on typical COM excitation temperatures of $\sim$\,5\,--\,6\,K, reported for the L1544 dust peak, the upper limit estimate for $T_\mathrm{ex} = 5$\,K might be better constrained at that position. It is on the same order than our \GlyOneS upper limit towards the B5 hotspot. At $T_\mathrm{ex} = 10$\,K, the upper limit at the L1544 dust peak becomes an order of magnitude lower than the limit at the B5 hotspot. Typical COM excitation temperatures at the L1544 methanol peak are reported as $\sim$\,5\,--\,8\,K. At both excitation limits ($T_\mathrm{ex} = 5$\,K and $T_\mathrm{ex} = 10$\,K), the \GlyOneS abundance upper limit towards the B5 hotspot is slightly lower but on the same order as at the L1544 methanol peak. We note that the conditions at the L1544 methanol peak are presumably more similar to the B5 methanol hotspot than the conditions at the L1544 dust peak.


 
Another 3-sigma upper limit of \GlyOneS is reported for the cold ($\sim$20\,K) outer layer of the solar-type protostar IRAS16293-2422 in \citet{Ceccarelli2000} as $5\times10^{12}$\,\si{\per\square\centi\meter}, i.e. $1\times10^{-10}$ w.r.t. $N(\ce{H2})$. The upper limit is obtained by averaging over three lines at around 101\,GHz. However, the assumed excitation conditions in the envelope of IRAS16293-2422, though cold, are only marginally comparable to the case of the B5 hotspot because a difference of $\geq$\,10\,K in excitation temperature is quite significant at low temperatures, and can change an upper limit by about an order of magnitude. 

\subsection{Glycine Upper Limits compared to COM Abundances at the B5 Hotspot}

The acetaldehyde column density we derive in our study, i.e. $(2.0\pm0.6)\times10^{12}$\,\si{\per\square\centi\meter}, is approximately 2.5 times lower than the estimate in \citet{Taquet2017}, i.e. $(5.2\pm0.7)\times10^{12}$\,\si{\per\square\centi\meter}. This discrepancy cannot be explained by the higher number of available data points for the rotation diagram analysis, but we are confident about the updated estimate.  

The abundances of COMs relative to \ce{H2} at the hotspot range from $2\times10^{-10}$ for acetaldehyde and $4\times10^{-10}$ for methyl formate to $2\times10^{-8}$ for methanol; di-methyl ether is tentatively detected at $2\times10^{-10}$ \citep{Taquet2017}. The most stringent, total abundance upper limit of glycine derived from the present observations ($2\times10^{-10}$ for $T_\mathrm{ex} = 7.5$\,K) is comparable to the abundance of the least abundant COM so far detected at the hotspot. Hence, given the present data, we can neither confirm nor rule out the presence of glycine in the gas phase at the B5 hotspot. If glycine is present but undetected, it is either less abundant than other COMs or has a more compact distribution or both.

\begin{table*}
\caption{\label{tab_ULcomparison}Glycine upper limits (3-sigma) towards cold molecular cloud sources.}
\centering
\begin{tabular}{cccccc}
\hline\hline
Source 	& Conformer 	& $T_\mathrm{ex}$ 	& $N_\mathrm{upper}$ 		&  $N_\mathrm{upper}/N(\ce{H2})$ 	& Ref.  \rule{0pt}{2.5ex} \\
		&			& [K] 				& [\si{\per\square\centi\meter}]	&  						& 				\\

\hline

L1544 Dust Peak	&	\GlyOne	&	5	&	$5.8\times10^{12}$	&	$1\times10^{-10}$	&	(1)	\rule{0pt}{2.5ex}\\
	&		&	10	&	$3.3\times10^{12}$	&	$6\times10^{-11}$	&		\\
\\											
L1544 Methanol Peak	&	\GlyOne	&	5	&	$9.5\times10^{12}$	&	$6\times10^{-10}$	&	(1)	\\
	&		&	10	&	$4.2\times10^{12}$	&	$3\times10^{-10}$	&		\\
\\											
IRAS16293-2422	&	\GlyOne	&	20	&	$5.0\times10^{12}$	&	$1\times10^{-10}$	&	(2)	\\
\\											
L134	&	\GlyTwo	&	10	&	$2.0\times10^{12}$	&	$2\times10^{-10}$	&	(3), (4)$^\mathrm{(a)}$	\\
\\											
TMC-1	&	\GlyTwo	&	7	&	$3.0\times10^{12}$	&	$2\times10^{-10}$	&	(5), (6)$^\mathrm{(a)}$ 	\\
\\											
L183	&	\GlyTwo	&	7	&	$4.5\times10^{12}$	&	$2\times10^{-10}$	&	(5), (7)$^\mathrm{(a)}$	\\

\hline

\multicolumn{6}{l}{\small{(1) \citet{JimenezSerra2016}; (2) \citet{Ceccarelli2000}; (3) \citet{Snyder1983};}} \rule{0pt}{2.5ex} \\
\multicolumn{6}{l}{\small{(4) \citet{vanderWerf1988}; (5) \citet{Brown1979}; (6) \citet{Spezzano2022};}} \\
\multicolumn{6}{l}{\small{(7) \citet{Lattanzi2020}}} \\
\multicolumn{6}{l}{$^\mathrm{(a)}$\small{\,reference for source \ce{H2} column density.}} \\

\end{tabular}
\end{table*}

\section{Conclusions}\label{sec_Conclusions}

Using the Onsala 20\,m telescope, we performed a deep search for glycine towards the B5 methanol hotspot, which is a cold source ($T_\mathrm{k} \approx 7.5$\,K) with yet a significant amount of COMs in the gas phase \citep{Taquet2017}. We targeted several transitions of \GlyOneS and \GlyTwoS in the frequency range $\sim$$70$\,--\,$78$\,GHz. Our study is the first to search for a set of glycine transitions in this lower frequency range. 

We did neither detect \GlyOneS nor \GlyTwoS during our observations but derive sensitive upper limits for both conformers, using spectral line stacking and assuming LTE conditions and optically thin emission. Since we did not detect glycine the hypothesis of a non-equilibrium \GlyTwo/\GlyOneS ratio similar to the case of $c$-\ce{HCOOH}/$t$-\ce{HCOOH} could not be tested. Our \GlyTwoS upper limits towards the B5 hotspot are the most stringent obtained so far for a cold molecular cloud source. The obtained \GlyOneS limits are mostly on the same order as previously published limits towards comparable sources. 

If glycine is present but undetected in the gas phase at the B5 hotspot it is either less abundant than other detected COMs in that source or has a more compact distribution or a combination thereof. We therefore do not rule out a future detection of glycine towards the B5 hotspot with higher-sensitivity instrumentation. \citet{JimenezSerra2014} calculate a detection limit of $\sim$\,$1.5\times10^{-11}$ for glycine in cold molecular cloud sources. This limit has not yet been reached by any study searching for glycine. 

\section*{Acknowledgements}

The authors acknowledge support from Onsala Space Observatory for the provisioning of its facilities/observational support. The Onsala Space Observatory national research infrastructure is funded through Swedish Research Council grant No 2017-00648. This work was supported by Chalmers Gender Initiative for Excellence (Genie). The work of SBC was supported by the Goddard Center for Astrobiology. This research has made use of data from the Herschel Gould Belt survey (HGBS) project (\url{http://gouldbelt-herschel.cea.fr}). The HGBS is a Herschel Key Programme jointly carried out by SPIRE Specialist Astronomy Group 3 (SAG 3), scientists of several institutes in the PACS Consortium (CEA Saclay, INAF-IFSI Rome and INAF-Arcetri, KU Leuven, MPIA Heidelberg), and scientists of the Herschel Science Center (HSC). We are grateful to Jose Aponte for preparing Figure 1, to Henrik Olofsson for assistance with the observations and discussions regarding the data reduction, and to the anonymous reviewer whose comments helped to improve the content of the paper.

\section*{Data Availability}

The full reduced spectrum as shown in Fig.\,\ref{fig_Spectrum} is available for download in table format.
 



\bibliographystyle{mnras}
\bibliography{Literature} 




\appendix

\section{Additional Figures}

\begin{figure*}
\resizebox{\hsize}{!}
{\includegraphics[width=12cm,clip]{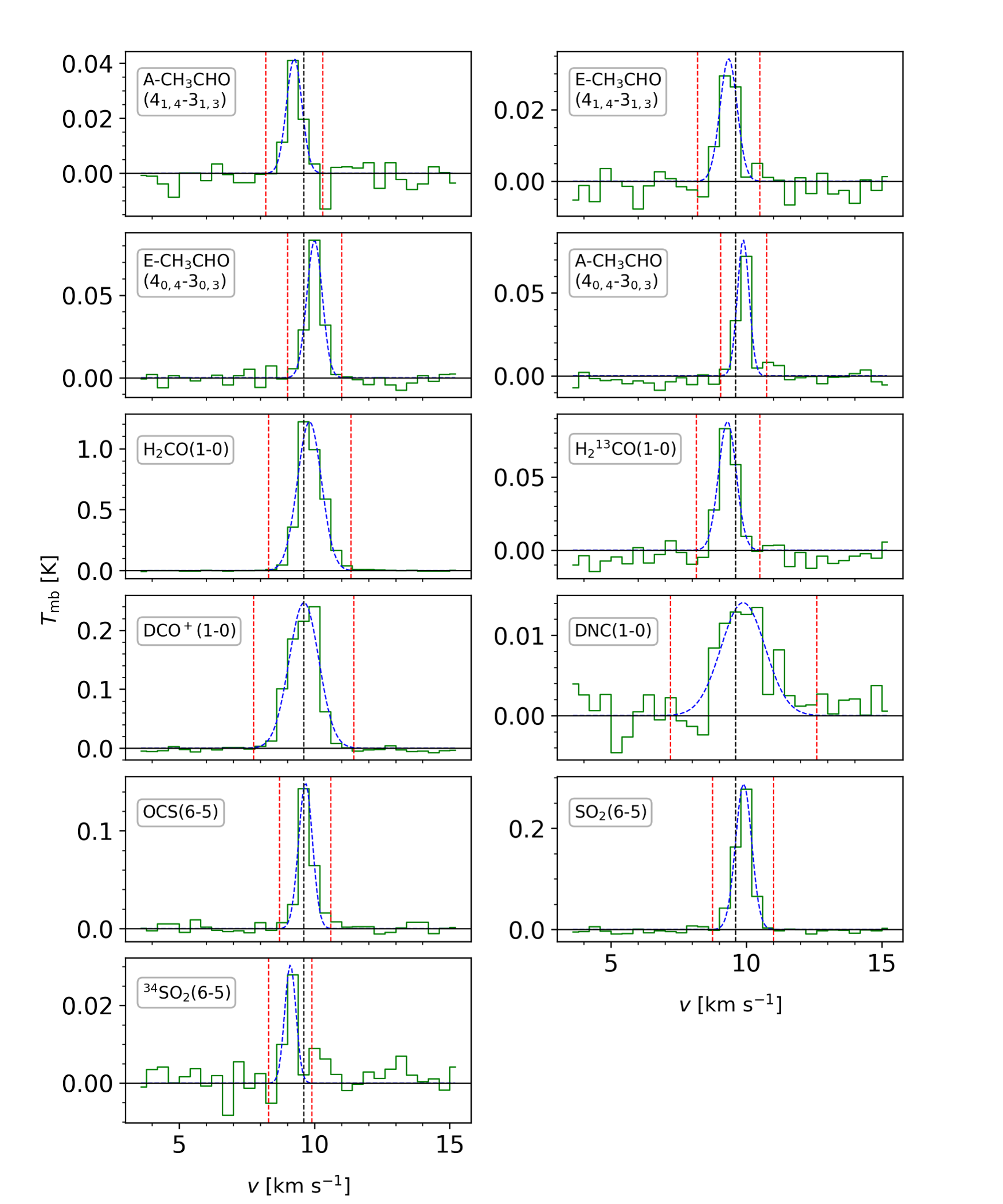}}
\caption{Spectral lines of non-targeted molecular transitions observed in the frequency range $70.2$\,-\,$77.9$\,GHz towards the B5 methanol hotspot. In all panels, the dashed blue line shows the Gaussian fit of the spectral line, and the dashed black line marks the LSR velocity of the hotspot. The dashed red lines mark the boundaries for the integrated intensity calculation.}
\label{fig_Spectra}
\end{figure*}

\begin{figure*}
\resizebox{\hsize}{!}
{\includegraphics[width=12cm,clip]{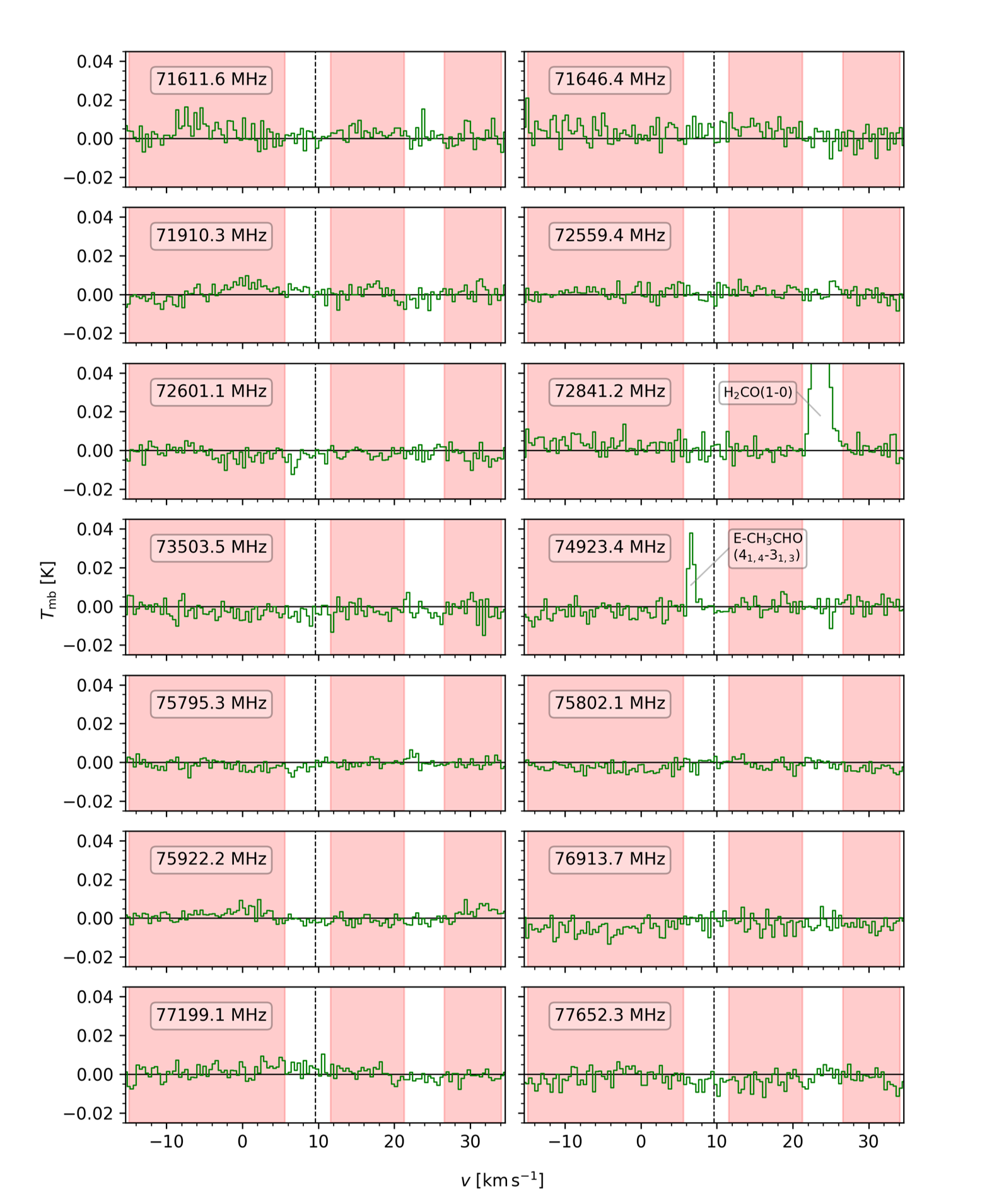}}
\caption{Spectra around transition frequencies of targeted \GlyOneS transitions. The red regions mark the ranges for the calculation of the RMS noise temperature. The dashed black line marks the LSR velocity of the B5 methanol hotspot.}
\label{fig_Gly1Spectra}
\end{figure*}

\begin{figure*}
\resizebox{\hsize}{!}
{\includegraphics[width=12cm,clip]{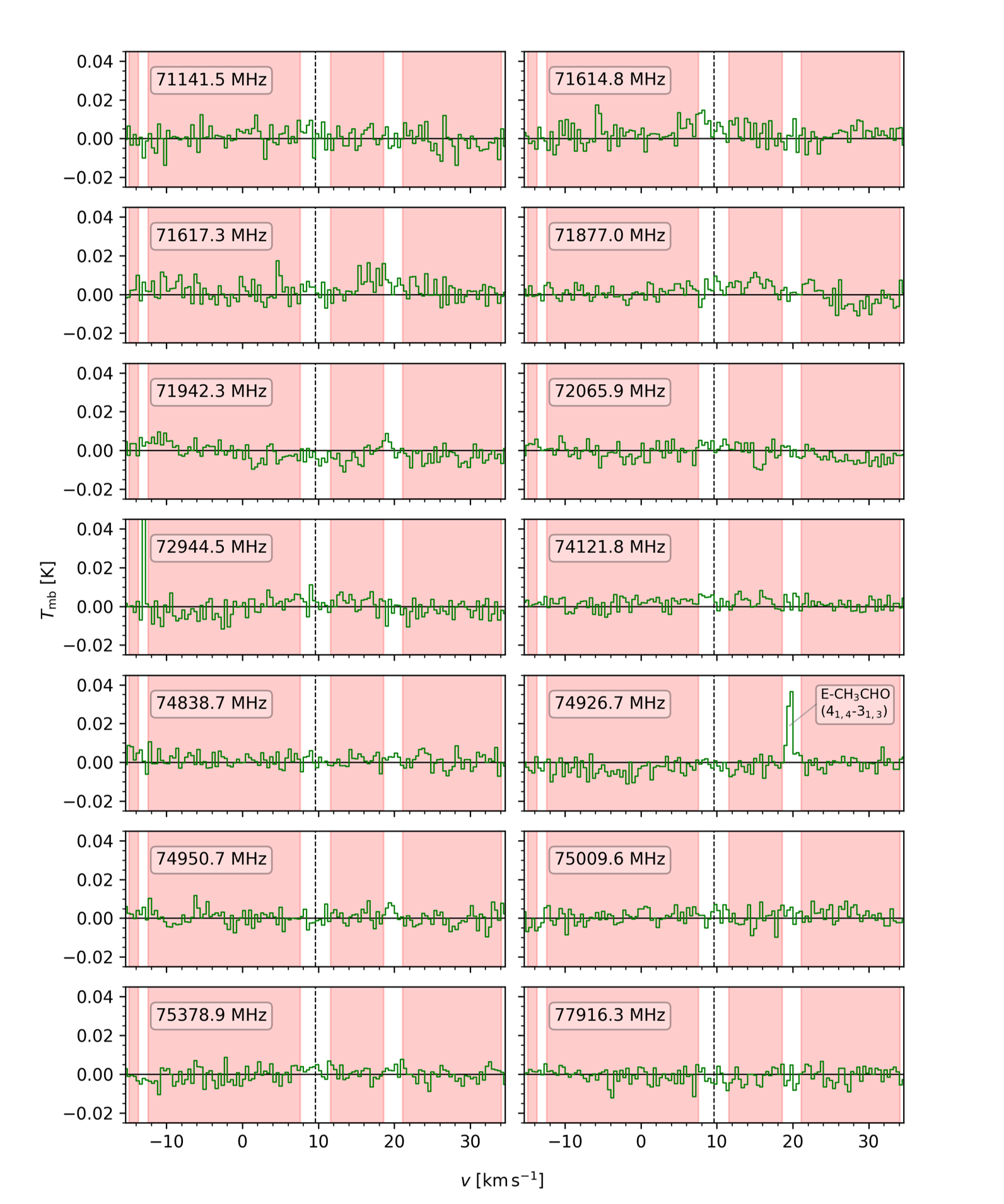}}
\caption{Spectra around transition frequencies of targeted \GlyTwoS transitions. The red regions mark the ranges for the calculation of the RMS noise temperature. The dashed black line marks the LSR velocity of the B5 methanol hotspot.}
\label{fig_Gly2Spectra}
\end{figure*}


\bsp	
\label{lastpage}
\end{document}